\documentclass[11pt]{article}%
\usepackage[onehalfspacing]{setspace}
\usepackage[left=1.1in,right=1.1in,top=1.1in,bottom=1.1in]{geometry}
\usepackage[colorlinks=true,linkcolor=RawSienna,citecolor=RawSienna,urlcolor=RawSienna,bookmarksopen=true,pdfstartview=FitB]%
{hyperref}
\usepackage{multicol}
\usepackage{natbib}
\usepackage{amsmath}
\usepackage{amsfonts}
\usepackage{amssymb}
\usepackage{graphicx}
\usepackage[dvipsnames]{xcolor}
\usepackage{float}
\usepackage{algorithm}
\usepackage{algpseudocode}
\usepackage{hyphenat}%
\providecommand{\U}[1]{\protect\rule{.1in}{.1in}}
\newtheorem{theorem}{Theorem}

\newtheorem{definition}{Definition}

\begin{document}

\title{\vspace{-1.4cm}False discovery rate control for multiple testing based on
p-values with c\`{a}dl\`{a}g distribution functions }
\author{Xiongzhi Chen\thanks{Corresponding author: Department of Mathematics and
Statistics, Washington State University, Pullman, WA 99164, USA; Email:
\texttt{xiongzhi.chen@wsu.edu}.} }
\date{}
\maketitle

\begin{abstract}
For multiple testing based on p-values with c\`{a}dl\`{a}g distribution
functions, we propose an FDR procedure ``BH+'' with proven conservativeness.
BH+ is at least as powerful as the BH procedure when they are applied to
super-uniform p-values. Further, when applied to mid p-values, BH+ is more
powerful than it is applied to conventional p-values. An easily verifiable
necessary and sufficient condition for this is provided. BH+ is perhaps the first
conservative FDR procedure applicable to mid p-values. BH+ is applied to
multiple testing based on discrete p-values in a methylation study, an HIV
study and a clinical safety study, where it makes considerably more
discoveries than the BH procedure.

\medskip


\end{abstract}

\section{Introduction}

\label{secIntro}

Multiple testing aiming at false discovery rate (FDR) control has been
routinely conducted in genomics, genetics and drug safety study, where test
statistics and their p-values are discrete. The discontinuities in p-value
distributions prevent popular FDR procedures, including the BH procedure
(``BH'') in \cite{Benjamini:1995} and Storey's procedure in \cite{Storey:2004}%
, to exhaust a target FDR bound to reach full power. This has motivated much
research in improving existing methods and developing new ones for multiple
testing with discrete statistics from three perspectives that are briefly
reviewed by \cite{Chen:2016c}. For example, \cite{Chen:2016c} proposed an
estimator of the proportion $\pi_{0}$ of true null hypotheses that is less
upwardly biased than that in \cite{Storey:2004} and an adaptive BH procedure
based on the estimator. However, the improvement of an adaptive procedure may
be very small when $\pi_{0}$ is very close to $1$. On the other hand, the
procedure in \cite{Liang:2015} is for discrete p-values with identical
distributions, the BHH procedure of \cite{Heyse:2011}, even though very
powerful, does not have implicit step-up critical constants and is hard to
analyze, and the procedures of \cite{Dohler:2017} can be more conservative
than the BH procedure.

Among all the methods just mentioned, each has been proposed for super-uniform
p-values, and none actively utilizes the super-uniformity of discrete p-values
or is able to deal with p-value with general distributions. On the other hand,
a mid p-value, which was proposed by \cite{Lancaster1961} and whose optimality
properties have been studied by \cite{Hwang:2001}, is almost surely smaller
than conventional p-values. So, a conservative FDR procedure based on mid
p-values may be more powerful than one based on conventional p-values, when
both are implemented with the same rejection constants. However, mid p-values
are sub-uniform (see \autoref{Def1} in \autoref{secDefinePval}), and there
does not seem to have been an FDR procedure that is based on mid p-values and
whose conservativeness is theoretically proven. This motivates us to
investigate FDR procedures that are applicable to p-values with general
distributions and effectively utilize the stochastic dominance, if any, of
their distributions with respect to the uniform distribution.

In this article, we deal with multiple testing based on p-values whose
cumulative distribution functions (CDF's) are right-continuous with
left-limits (i.e., c\`{a}dl\`{a}g). We propose a new FDR procedure, referred
to as \textquotedblleft BH+\textquotedblright, that is conservative when
p-values satisfy the property of \textquotedblleft positive regression
dependency on subsets of the set of true null hypotheses (PRDS)'' and have
c\`{a}dl\`{a}g CDF's. BH+ generalizes BH for multiple testing based on
p-values. For multiple testing with discrete, super-uniform p-values under
PRDS, BH+ is at least as powerful as BH. Further, BH+ can be applied to mid
p-values and is more powerful than BH in this setting. In contrast, BH has not
been theoretically proven to be conservative when it is applied to mid
p-values. An easy-to-check necessary and sufficient condition is given on when
BH+ based on mid p-values rejects at least as many null hypotheses as it does
based on conventional p-values. BH+ is perhaps the first conservative FDR
procedure that is applicable to sub-uniform p-values, and is implemented by the \textsf{R} package
``\href{https://cran.r-project.org/web/packages/fdrDiscreteNull/index.html}{fdrDiscreteNull}%
'' available on \href{https://cran.r-project.org/}{CRAN}.

The rest of the article is organized as follows. \autoref{secNew} discusses
two definitions of a two-sided p-value for a discrete statistic and
introduces the BH+ procedure. \autoref{secSim} contains a simulation study on
BH+ when they are applied to p-values of Binomial tests (BT's) and Fisher's
exact tests (FET's). \autoref{seAppAll} provides three applications of BH+.
The article ends with a discussion in \autoref{secDisc}. All
proofs and additional simulation results are relegated into the appendices.

\section{Multiple testing with p-values and the BH+ procedure}

\label{secNew}

We will collect in \autoref{secDefinePval} two definitions of a two-sided
p-value and introduce the BH+ procedure in \autoref{secNonAdap}.

\subsection{Definitions of a two-sided p-value}

\label{secDefinePval}

For a random variable $X$, let $F$ be its CDF with support $\mathcal{S}$ and
probability density function (PDF) $f$. Here $f$ is the Radon-Nikodym
derivative $\frac{\mathsf{d}F}{\mathsf{d}\upsilon}$, where $\upsilon$ is the
Lebesgue measure or the counting measure on $\mathcal{S}$. For an observation
$x_{0}$ from $X$, set%
\[
l\left(  x_{0}\right)  =\int_{\left\{  x\in\mathcal{S}:f\left(  x\right)
<f\left(  x_{0}\right)  \right\}  }\mathsf{d}F\left(  x\right)  \text{ \ and
\ }e\left(  x_{0}\right)  =\int_{\left\{  x\in\mathcal{S}:f\left(  x\right)
=f\left(  x_{0}\right)  \right\}  }\mathsf{d}F\left(  x\right)  .
\]
Following \cite{Agresti:2002}, let $P\left(  x_{0}\right)  =l\left(
x_{0}\right)  +e\left(  x_{0}\right)  $ be the conventional two-sided p-value
of $x_{0}$, and following \cite{Hwang:2001}, let $Q\left(  x_{0}\right)
=l\left(  x_{0}\right)  +2^{-1}e\left(  x_{0}\right)  $ be the two-sided mid p-value.

\begin{definition}
\label{Def1}A random variable $X$ with range $\left[  0,1\right]  $ is called
\textquotedblleft super-uniform\textquotedblright\ if $\Pr\left(  X\leq
t\right)  \leq t$ for all $t\in\left[  0,1\right]  $, and it is called
\textquotedblleft sub-uniform\textquotedblright\ if $\Pr\left(  X\leq
t\right)  \geq t$ for all $t\in\left[  0,1\right]  $.
\end{definition}

$P$ is super-uniform and $\Pr\left(  P\left(  X\right)  \leq P\left(
x\right)  \right)  =P\left(  x\right)  $ for all $x\in\mathcal{S}$, whereas
$Q$ is sub-uniform and $\Pr\left(  Q\left(  X\right)  \leq Q\left(  x\right)
\right)  =P\left(  x\right)  $ for all $x\in\mathcal{S}$.

\subsection{The BH+ procedure}

\label{secNonAdap}

Consider simultaneously testing $m$ null hypotheses $\left\{  H_{i}\right\}
_{i=1}^{m}$ based on their associated p-values $\left\{  p_{i}\right\}
_{i=1}^{m}$ with matching indices. Among $\left\{  H_{i}\right\}  _{i=1}^{m}$,
$m_{0}$ are true and the rest $m_{1}$ false. Let $I_{0}$ be the index set of
true null hypotheses, then $m_{0}$ is the cardinality of $I_{0}$ and $\pi_0 = m_0 m^{-1}$.
For each $i$, let $F_{i}$ be the CDF of $p_{i}$ obtained by assuming
$H_{i}$ is a true null, and assume $F_{i}$ has a finite support $S_{i}$. We do
not require a p-value to be super-uniform under the null hypothesis but do
assume that $F_{i}\left(  s\right)  =s$ for $s\in S_{i}$ when a $p_{i}$ is
super-uniform. Let $\left\{  p_{\left(  i\right)  }\right\}  _{i=1}^{m}$ be
the order statistics of $\left\{  p_{i}\right\}  _{i=1}^{m}$ such that
$p_{\left(  1\right)  }\leq p_{\left(  2\right)  }\leq\cdots\leq p_{\left(
m\right)  }$. Further, let $H_{\left(  i\right)  }$ be the null hypothesis
associated with $p_{\left(  i\right)  }$ for each $i$. The BH+ procedure is
given by \autoref{Alg:uBHProc}.

\begin{algorithm}[h!]
\begin{algorithmic}[1]
\caption{The BH+ procedure}\label{Alg:uBHProc}
\State 
Let $F^{\ast}   =\max_{1\leq i\leq m}F_{i} $
and denote its support by $S^{\ast}$.
\State 
At nominal FDR level $\alpha \in \left(0,1\right)$, define critical values%
\begin{equation}
\gamma_{k}=\max\left\{  t\in S^{\ast}:F^{\ast}\left(  t\right)  \leq\frac{\alpha
k}{m}\right\}  \text{ \ for \ }k=1,\ldots,m.\label{eqcca}%
\end{equation}
\State 
Let $\mathcal{D} = \left\{  1\leq i\leq m:p_{\left(  k\right)  }\leq\gamma
_{k}\right\}$. If $\mathcal{D} = \varnothing$, reject no null hypotheses; otherwise,
set $R=\max \mathcal{D}$ and reject each $H_{i}$ if its associated
p-value $p_{i}\leq\gamma_{R}$.
\end{algorithmic}
\end{algorithm}{}

When $\left\{  p_{i}\right\}  _{i=1}^{m}$ are discrete and super-uniform,
$F^{\ast}\left(  t\right)  \leq t$ for each $t\in\left[  0,1\right]  $ and
elementwise $\left\{  \gamma_{k}\right\}  _{k=1}^{m}$ are no smaller than the
critical values of the BH procedure (\textquotedblleft BH\textquotedblright%
\ for short). So, BH+ directly utilizes the super-uniformity of p-values to
obtain potentially larger critical values than those of BH, and
the more $F^{\ast}$ deviates from the identity function $\iota$ when both are
restricted on $\left[  0,\gamma_{m}\right]  $, the more BH+ improves on BH.
BH+ reduces to BH when $\left\{  p_{i}\right\}  _{i=1}^{m}$
are all $\mathsf{Unif}\left(  0,1\right)  $ under the null hypotheses but it
does not when some of $\left\{  p_{i}\right\}  _{i=1}^{m}$ are discrete.

The conservativeness of the BH+ procedure is justified by:

\begin{theorem}
\label{lmGeneral}Let $R$ be the number of rejections made by the BH+
procedure. If\ $\left\{  p_{i}\right\}  _{i=1}^{m}$ satisfy \textquotedblleft
positive regression dependency on subsets of the set of true null hypotheses
(PRDS)\textquotedblright, i.e.,%
\begin{equation}
\Pr\left(  \left.  R\geq r\right\vert p_{i}\leq t\right)  \text{ is
non-increasing in }t\in\left(  0,\gamma_{r}\right]  \text{ for each }i\in
I_{0}\text{ and }r\in\left\{  1,\ldots m\right\}  , \label{eqPRDS}%
\end{equation}
then the BH+ procedure is conservative. If further $\left\{  p_{i}\right\}
_{i=1}^{m}$ are super-uniform, then almost surely the set of null hypotheses
rejected by the BH+ procedure contains that of the BH procedure.
\end{theorem}

The definition of PRDS stated in (\ref{eqPRDS}) is borrowed from condition
(D1) in Section 4 of \cite{Finner:2009}, and the conservativeness of the BH+
procedure under PRDS does not require the super-uniformity of p-values. For
example, BH+ can be applied to two-sided mid p-values and is in this case
usually more powerful than it is applied to conventional p-values; see
simulation study in \autoref{secSim}. Since PRDS
holds for independent p-values, BH+ is conservative under independence.

Our next result shows when the BH+ procedure based on two-sided mid p-values
rejects at least as many null hypotheses as it does based on two-sided
conventional p-values. In order to state it, we need some additional
notations. For each $i$ and null hypothesis $H_{i}$, let $P_{i}$ be its
associated two-sided conventional p-value with CDF $F_{i}^{\mathsf{cp}}$, and
$Q_{i}$ the corresponding two-sided mid p-value with CDF $F_{i}^{\mathsf{mp}}%
$. Let $W_{\mathsf{mp}}=\max_{1\leq i\leq m}F_{i}^{\mathsf{mp}}$ and
$W_{\mathsf{cp}}=\max_{1\leq i\leq m}F_{i}^{\mathsf{cp}}$. Further, let
$\left\{  P_{\left(  i\right)  }\right\}  _{i=1}^{m}$ be the order statistics
of $\left\{  P_{i}\right\}  _{i=1}^{m}$ and $\left\{  Q_{\left(  i\right)
}\right\}  _{i=1}^{m}$ those of $\left\{  Q_{i}\right\}  _{i=1}^{m}$.

\begin{theorem}
\label{ThmMidVersusCon}$W_{\mathsf{mp}}\left(  t\right)  \geq W_{\mathsf{cp}%
}\left(  t\right)  $ for all $t\in\left[  0,1\right]  $. At the same nominal
FDR level $\alpha\in\left(  0,1\right)  $, let $R_{\mathsf{cp}}$ be the number
of rejected null hypotheses of BH+ based on $\left\{  P_{i}\right\}
_{i=1}^{m}$ and $R_{\mathsf{mp}}$ that of BH+ based on $\left\{
Q_{i}\right\}  _{i=1}^{m}$. Then $R_{\mathsf{mp}}\geq R_{\mathsf{cp}}$ if and
only if%
\begin{equation}
W_{\mathsf{mp}}\left(  Q_{\left(  R_{\mathsf{cp}}\right)  }\right)  \leq\alpha
m^{-1}R_{\mathsf{cp}}. \label{eq15a}%
\end{equation}

\end{theorem}

Condition (\ref{eq15a}) is easy to check. Let $\mathbb{F}_{m}^{\mathsf{cp}}$
and $\mathbb{F}_{m}^{\mathsf{mp}}$ be the empirical CDF's of $\left\{
P_{i}\right\}  _{i=1}^{m}$ and $\left\{  Q_{i}\right\}  _{i=1}^{m}$
respectively. If we take \textquotedblleft time\textquotedblright\ as the
value of a p-value, then $P_{\left(  R_{\mathsf{cp}}\right)  }$ is the last
time right after which $\left\{  W_{\mathsf{cp}}\left(  t\right)  ,t\in\left[
0,1\right]  \right\}  $ crosses $\left\{  \alpha\mathbb{F}_{m}^{\mathsf{cp}%
}\left(  t\right)  :t\in\left[  0,1\right]  \right\}  $ from below, and
\autoref{ThmMidVersusCon} says that the last crossing from below of the level
$\alpha\mathbb{F}_{m}^{\mathsf{cp}}\left(  P_{\left(  R_{\mathsf{cp}}\right)
}\right)  $ by $\left\{  W_{\mathsf{mp}}\left(  t\right)  ,t\in\left[
0,1\right]  \right\}  $ should be later than $Q_{\left(  R_{\mathsf{cp}%
}\right)  }$ in order for the BH+ procedure based on $\left\{  Q_{i}\right\}
_{i=1}^{m}$ to reject as least as many null hypotheses as it does based on
$\left\{  P_{i}\right\}  _{i=1}^{m}$. This is very intuitive since
$W_{\mathsf{mp}}\left(  t\right)  \geq W_{\mathsf{cp}}\left(  t\right)  $ for
all $t\in\left[  0,1\right]  $. Note that \autoref{ThmMidVersusCon} poses no
restrictions on the dependence structures among $\left\{  Q_{i}\right\}
_{i=1}^{m}$ or $\left\{  P_{i}\right\}  _{i=1}^{m}$.

\section{Simulation study}

\label{secSim}

We now present a simulation study on BH+ based on two-sided p-values of
Binomial tests (BT's) and Fisher's exact tests (FET's). BH+ will be compared
to BH.

\subsection{Binomial test and Fisher's exact test}

\label{SecTwoTests}

The Binomial test (BT) is used to test if two independent Poisson random
variables $X_{i}\sim\mathsf{Poisson}\left(  \lambda_{i}\right)  ,i=1,2$ have
the same means $\lambda_{1}=\lambda_{2}$. Specifically, after a count $c_{i}$
is observed from $X_{i}$ for each $i\in\left\{  1,2\right\}  $, the BT
statistic $T_{\theta}\sim\mathsf{Binomial}\left(  \theta,c\right)  $ with
probability of success $\theta=\lambda_{1}\left(  \lambda_{1}+\lambda
_{2}\right)  ^{-1}$ and total number of trials $c=c_{1}+c_{2}$ as the observed
total count. Further, the p-value associated with $T_{\theta}$ is computed
using the CDF of $T_{0.5}$ under the null hypothesis $\theta=0.5$. Note that
the PDF of $X\sim\mathsf{Binomial}\left(  0.5,n\right)  $ with $n\in
\mathbb{N}$ is%
\begin{equation}
f\left(  x;n\right)  =\binom{n}{x}2^{-n}\text{ \ for }x=0,1,\ldots,n.
\label{eqc4}%
\end{equation}

On the other hand, Fisher's exact test (FET) has been widely used in assessing
if a discrete conditional distribution is identical to its unconditional
version, where the observations are modelled by Binomial distributions.
Specifically, after a count $c_{i}$ is observed from $X_{i}\sim
\mathsf{Binomial}\left(  q_{i},N_{i}\right)  $ for each $i\in\left\{
1,2\right\}  $, the marginal $\mathbf{N}=\left(  N_{1},N_{2},M\right)  $ is
formed with $M=c_{1}+c_{2}$ as the observed total count, and the test statistic
$T_{\theta}$ follows a hypergeometric distribution $\mathsf{HGeom}\left(
\theta,\mathbf{N}\right)  $ with $\theta=\frac{q_{1}\left(  1-q_{2}\right)
}{q_{2}\left(  1-q_{1}\right)  }$ for $q_{1},q_{2}\in\left(  0,1\right)  $.
The p-value associated with $T_{\theta}$\ for the observation $c_{1}$ is
defined using the CDF\ of $T_{1}$ under the null hypothesis $\theta=1$. If
$N_{1}=N_{2}$, then the distribution of $T_{1}$ only depends on $M$ and has
PDF%
\begin{equation}
g\left(  x;N,M\right)  =\left.  \binom{N}{x}\binom{N}{M-x}\right/  \binom
{2N}{M}\text{ for }0\leq x\,\leq M. \label{eqc6}%
\end{equation}

\subsection{Simulation design}

\label{SecSimDesign}

Set $m=200$, $\pi_{0}$ $=0.5,0.6,0.7,0.8$ or $0.9$, and $\alpha=0.05,0.1,0.15$
or $0.2$. Recall $m_{0}=m\pi_{0}$ and $m_{1}=m-m_{0}$. Independent Poisson and
Binomial data are generated follows:

\begin{itemize}
\item Poisson data: let $\mathsf{Pareto}\left(  \eta,\sigma\right)  $ denote
the Pareto distribution with location $\eta$ and shape $\sigma$. Generate $m$
$\theta_{i1}$'s independently from $\mathsf{Pareto}\left(  \eta,5\right)  $
with $\eta=3, 4.5$ or $6$. Generate $m_{1}$ $\rho_{i}$'s independently from
$\mathsf{Unif}\left(  3,5.5\right)  $. Set $\theta_{i2}=\theta_{i1}$ for
$1\leq i\leq m_{0}$, $\theta_{i2}=\rho_{i}\theta_{i1}$ for $m_{0}+1\leq i\leq
m_{0}+\left[  0.5m_{1}\right]  $, and $\theta_{i1}=\rho_{i}\theta_{i2}$ for
$m_{0}+\left[  0.5m_{1}\right]  +1\leq i\leq m$, where $\left[  x\right]  $ is
the integer part of $x\in\mathbb{R}$. For each $i$ and $j\in\left\{
1,2\right\}  $, independently generate a count $c_{ij}$ from $\mathsf{Poisson}%
\left(  \theta_{ij}\right)  $.

\item Binomial data: generate $\theta_{i1}$ from $\mathsf{Unif}\left(
0.2,0.3\right)  $ for $i=1,\ldots,m_{0}$ and set $\theta_{i2}=\theta_{i1}$ for
$i=1,\ldots,m_{0}$. Set $\theta_{i1}=0.3$ and $\theta_{i2}=0.75$ for
$m_{0}+1\leq i\leq m_{0}+\left[  0.5m_{1}\right]  $ but $\theta_{i1}=0.75$ and
$\theta_{i2}=0.3$ for $m_{0}+\left[  0.5m_{1}\right]  +1\leq i\leq m$. Set
$n=10$, $20$ or $30$, and for each $j\in\{1,2\}$ and $i$, independently
generate a count $c_{ij}$ from $\mathsf{Binomial}\left(  \theta_{ij},n\right)
$.
\end{itemize}

In contrast, positively and blockwise correlated Poisson and Binomial data are
generated as follows:

\begin{itemize}
\item Construct a block diagonal, correlation matrix $\mathbf{D}%
=\mathsf{diag}\left\{  \mathbf{D}_{1},\mathbf{D}_{2},\mathbf{D}_{3}%
,\mathbf{D}_{4},\mathbf{D}_{5}\right\}  $ with $5$ blocks, such that each
$\mathbf{D}_{i}$ is of size $40\times40$ and that the off-diagonal entries of
$\mathbf{D}_{i}$ are identically $0.2$. Generate a realization $\mathbf{z}%
=(z_{1},\ldots,z_{m})$ from the $m$-dimensional Normal distribution with zero
mean and correlation matrix $\mathbf{D}$, and obtain the vector $\mathbf{u}%
=(u_{1},\ldots,u_{m})$ such that $u_{i}=\Phi(z_{i})$, where $\Phi$ is the CDF
of the standard Normal random variable.

\item Maintain the same parameters used to generate independent Poisson and
Binomial data, and for each $j\in\left\{  1,2\right\}  $ and $i\in\left\{
1,\ldots,m\right\}  $, generated a count $c_{ij}$ corresponds to quantile
$u_{i}$ of the CDF of $\mathsf{Poisson}\left(  \theta_{ij}\right)  $ or
$\mathsf{Binomial}\left(  \theta_{ij},n\right)  $.
\end{itemize}

With $c_{ij}$, $j=1,2$ for each $i$, conduct the BT\ or FET on $H_{i0}%
:\theta_{i1}=\theta_{i2}\text{ versus }H_{i1}:\theta_{i1}\neq\theta_{i2}$ and
obtain the two-sided conventional or mid p-value $p_{i}$ for the test. Apply
the FDR procedures to the $m$ p-values $\left\{  p_{i}\right\}  _{i=1}^{m}$.
Each experiment is determined by a triple $\left(  \alpha,\pi_{0},\eta\right)
$ or $\left(  \alpha,\pi_{0},n\right)  $ and repeated $300$ times to obtain
statistics for the performances of the FDR procedures.

The signal strengths in the simulation design prevent a BT and FET to have
very lower power. In addition, to assess how the support of the maximal CDF of
p-values affects the performance of BH+, we have chosen $3$ Pareto
distributions with different location parameters to generate the means for
Poisson data, and Binomial distributions with $3$ different numbers of total
trials to generate Binomial data.

\subsection{Simulation results}

\label{secSimRes}

We use the expectation of the true discovery proportion (TDP), defined as the
ratio of the number of rejected false null hypotheses to the total number of
false null hypotheses, to measure the power of an FDR procedure. Recall that
the FDR is the expectation of the false discovery proportion (FDP).
We also report the standard deviations of the FDP
and TDP since smaller standard deviations for these quantities mean that the
corresponding procedure is more stable in FDR and power. For the simulations,
\textquotedblleft BH\textquotedblright\ is the BH procedure, \textquotedblleft
BH+\textquotedblright\ is BH+ applied to conventional p-values, and
\textquotedblleft MidPBH+\textquotedblright\ is BH+ applied to mid p-values.

\begin{figure}[h]
\centering
\includegraphics[height=0.84\textheight,width=\textwidth]{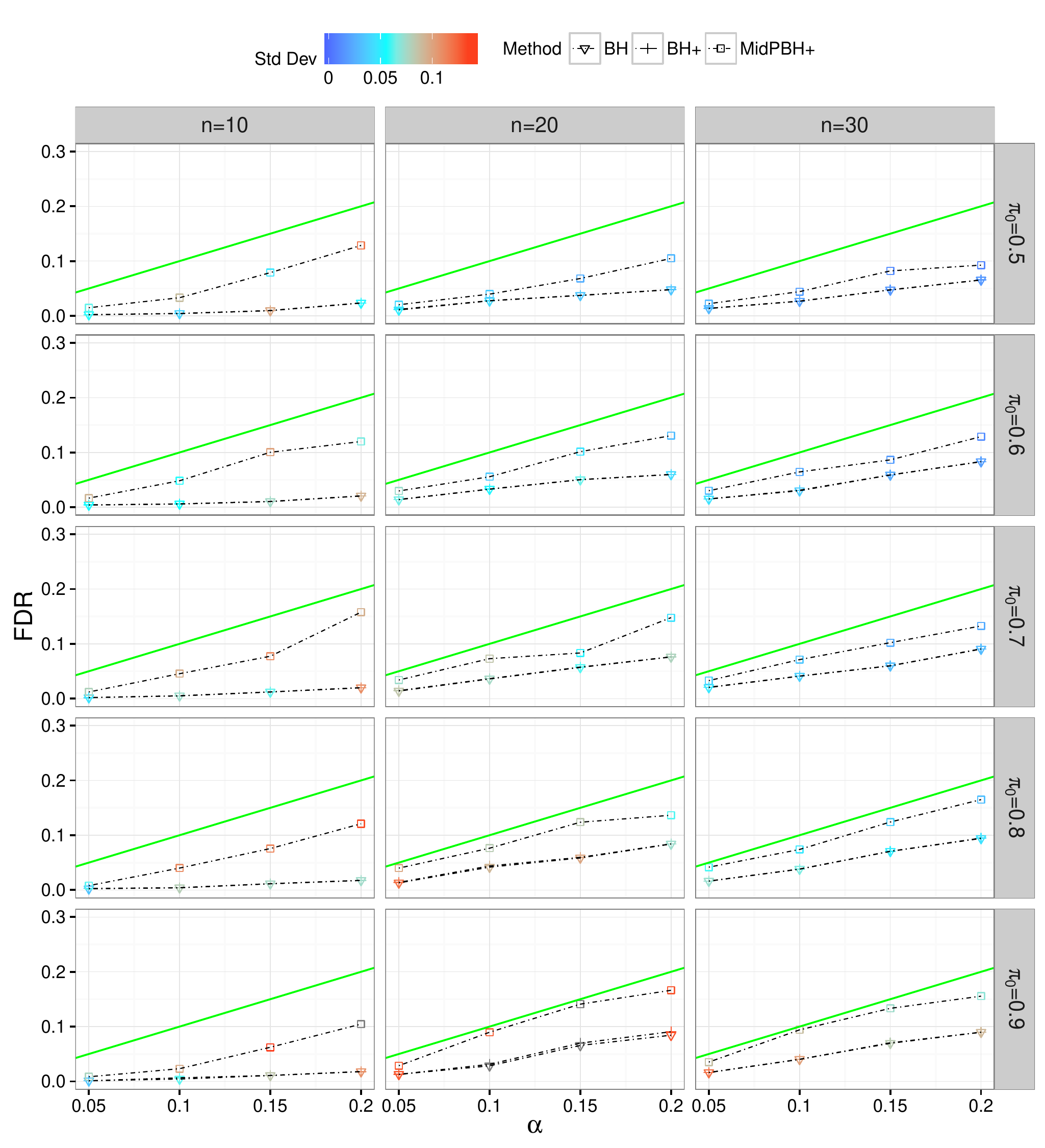}
\vspace{-0.5cm} \caption[FDR for FET]{False discovery rate (FDR) of each
procedure. The color legend ``Std Dev" denotes the standard deviation of the
false discovery proportion (FDP) whose expectation is FDR. The diagonal line
indicates equality of the nominal FDR level $\alpha$ and the FDR of a
procedure. Each procedure has been applied to two-sided p-values of Fisher's
exact tests under independence.}%
\label{figFdrFET}%
\end{figure}

\begin{figure}[h]
\centering
\includegraphics[height=0.84\textheight,width=\textwidth]{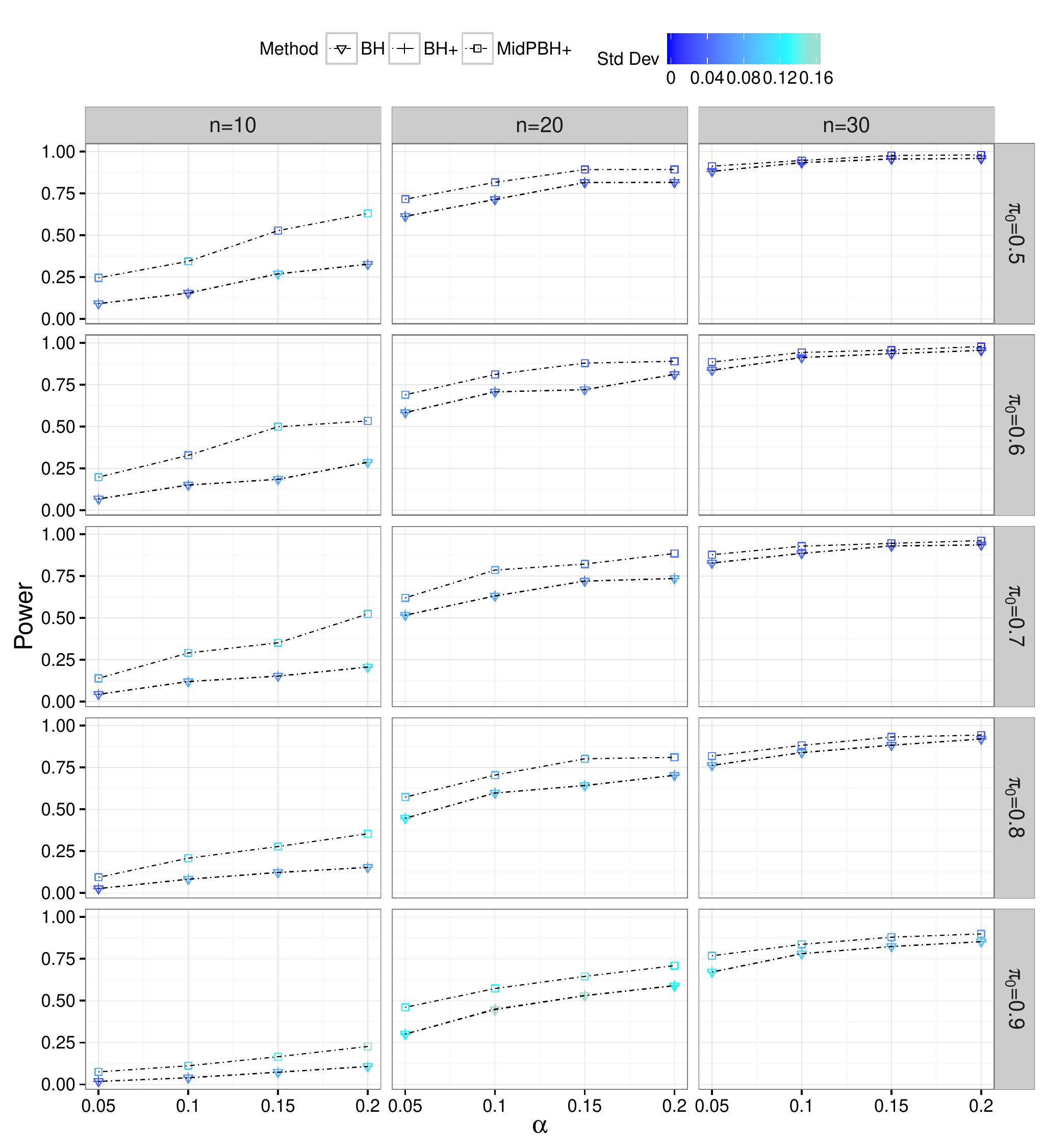}
\vspace{-0.5cm} \caption[Power for FET]{Power of each procedure. The color
legend ``Std Dev" denotes the standard deviation of the true discovery
proportion (TDP) whose expectation is power. Each procedure has been applied
to two-sided p-values of Fisher's exact tests under independence.}%
\label{figPwrFET}%
\end{figure}

\begin{figure}[h]
\centering
\includegraphics[height=0.84\textheight,width=\textwidth]{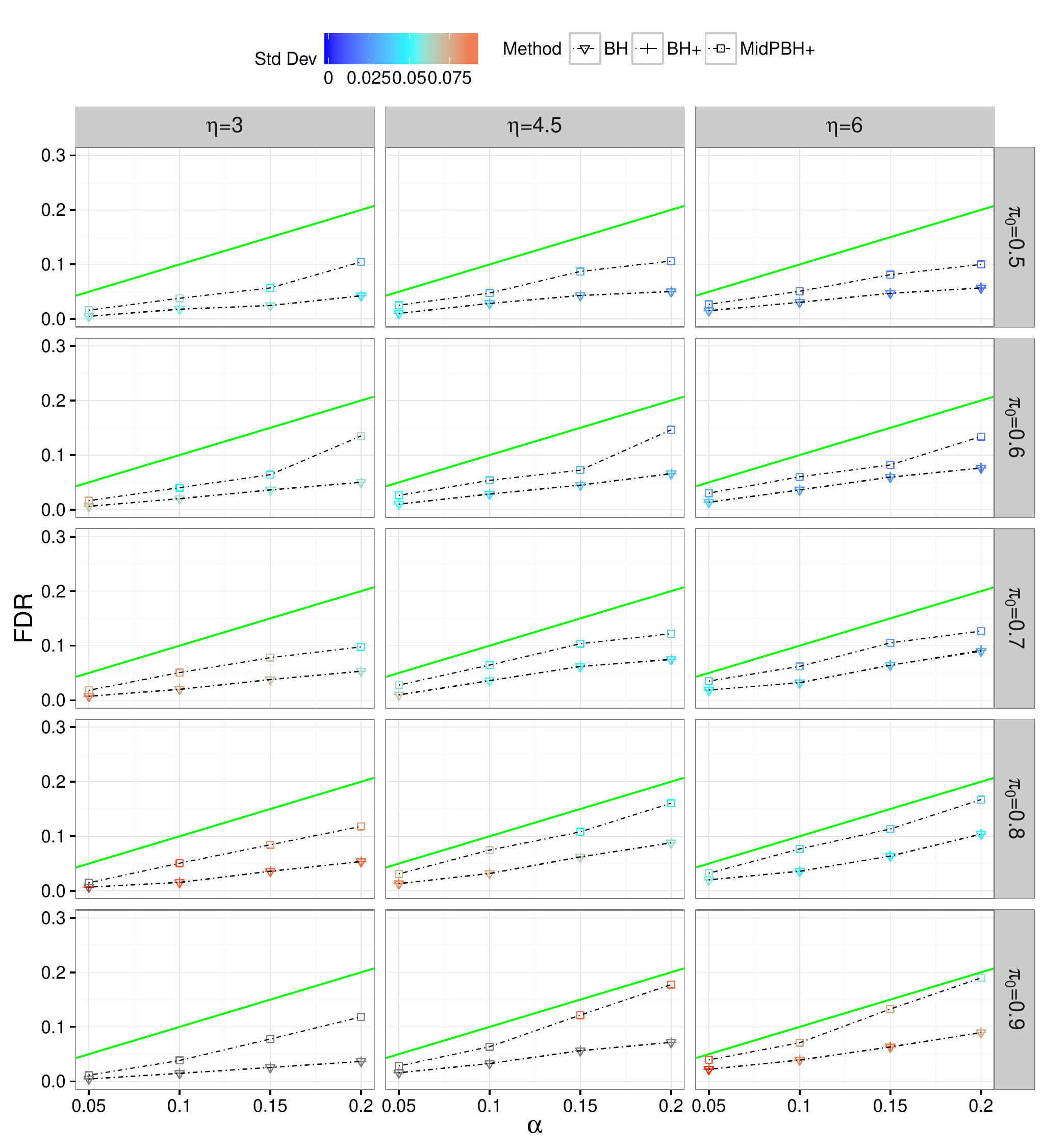}
\vspace{-0.5cm} \caption[FDR for BT]{False discovery rate (FDR) of each
procedure. The color legend ``Std Dev" denotes the standard deviation of the
false discovery proportion (FDP) whose expectation is FDR. The diagonal line
indicates equality of the nominal FDR level $\alpha$ and the FDR of a
procedure. Each procedure has been applied to two-sided p-values of Binomial
tests under independence.}%
\label{figFdrBT}%
\end{figure}

\begin{figure}[h]
\centering
\includegraphics[height=0.84\textheight,width=\textwidth]{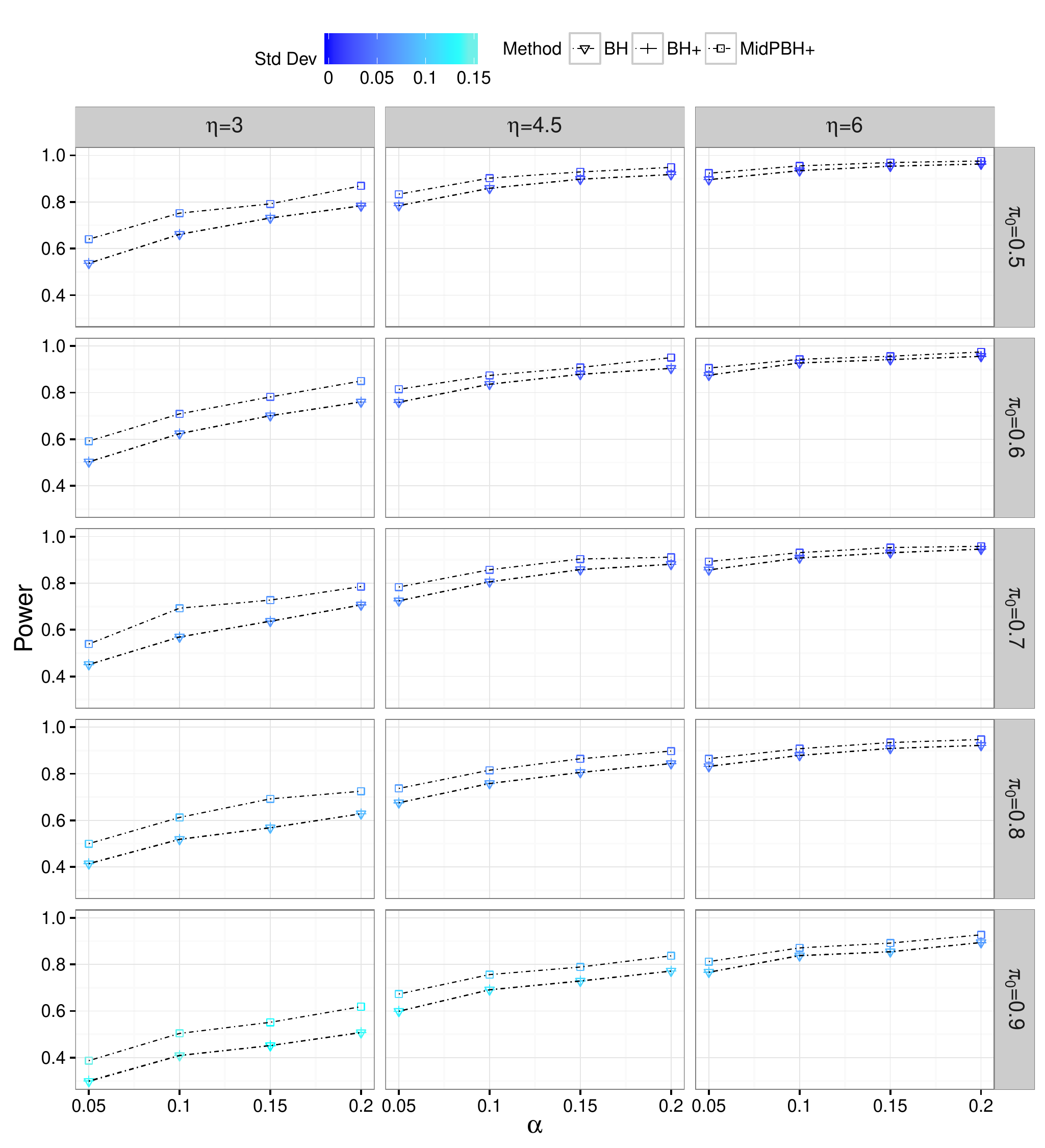}
\vspace{-0.5cm} \caption[Power for BT]{Power of each FDR procedure. The color
legend ``Std Dev" denotes the standard deviation of the true discovery
proportion (TDP) whose expectation is the power. Each procedure has been
applied to two-sided p-values of Binomial tests under independence.}%
\label{figPwrBT}%
\end{figure}

We first summarize the results under independence. \autoref{figFdrFET} and
\autoref{figFdrBT} record the FDR of each procedure. Both BH+ and MidPBH+ are
conservative. \autoref{figPwrFET} and \autoref{figPwrBT} record the power of
each procedure. BH+ is not less powerful than BH, and MidPBH+ is usually more
powerful than BH+ and BH. The improvement in power of MidPBH+ over BH+ can be
considerable for FET's when the total number of trials for Binomial
distributions is not large and for BT's when the means of Poisson
distributions are not large.

\begin{figure}[H]
\centering
\includegraphics[height=0.8\textheight,width=\textwidth]{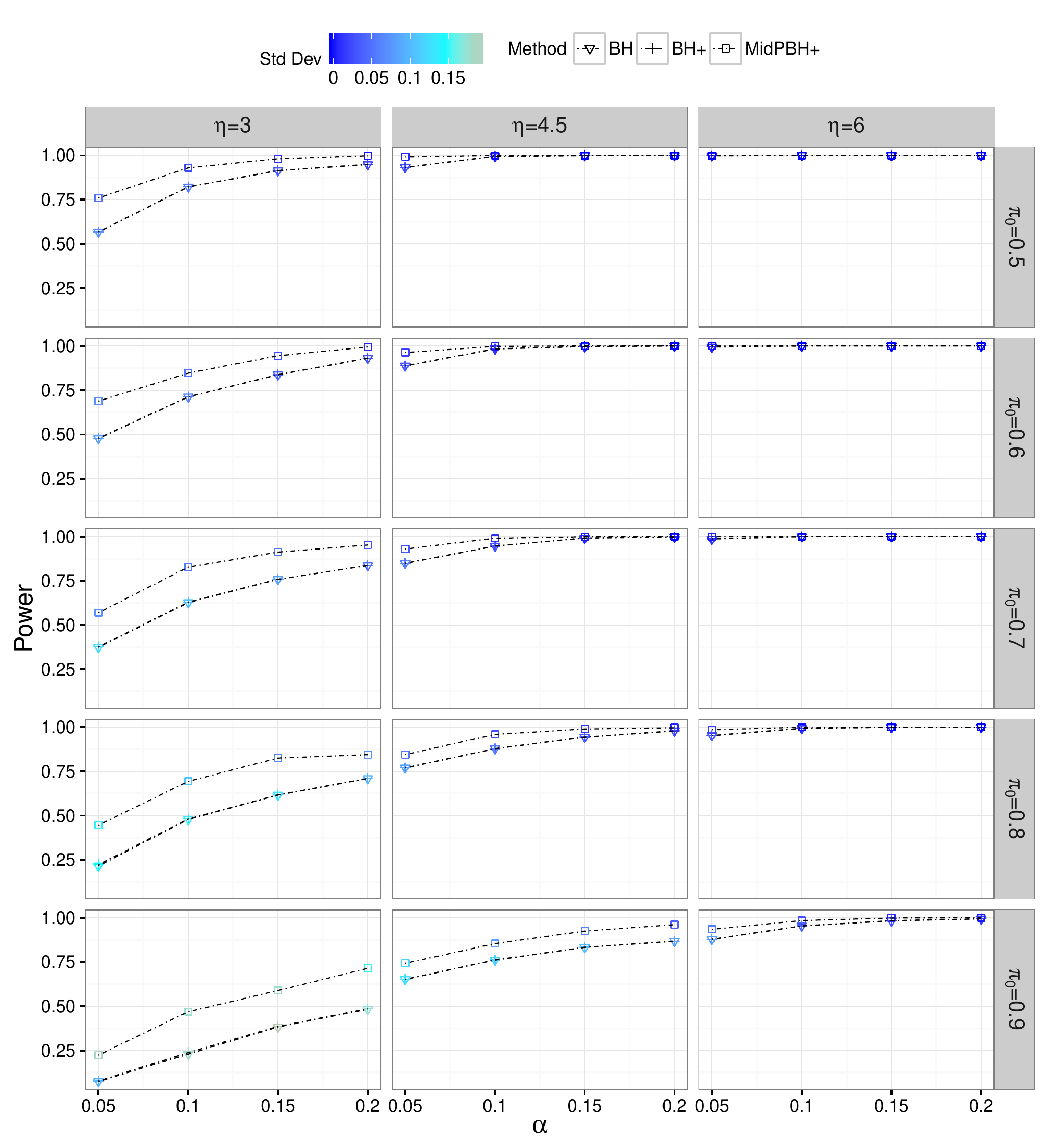}
\vspace{-0.5cm} \caption[Power for BT Dep]{Power of each procedure. The color
legend ``Std Dev" denotes the standard deviation of the true discovery
proportion (TDP) whose expectation is power. Each procedure has been applied
to two-sided p-values of Binomial tests under positive dependence.}%
\label{figPwrBTDep}%
\end{figure}

\begin{figure}[h]
\centering
\includegraphics[height=0.84\textheight,width=\textwidth]{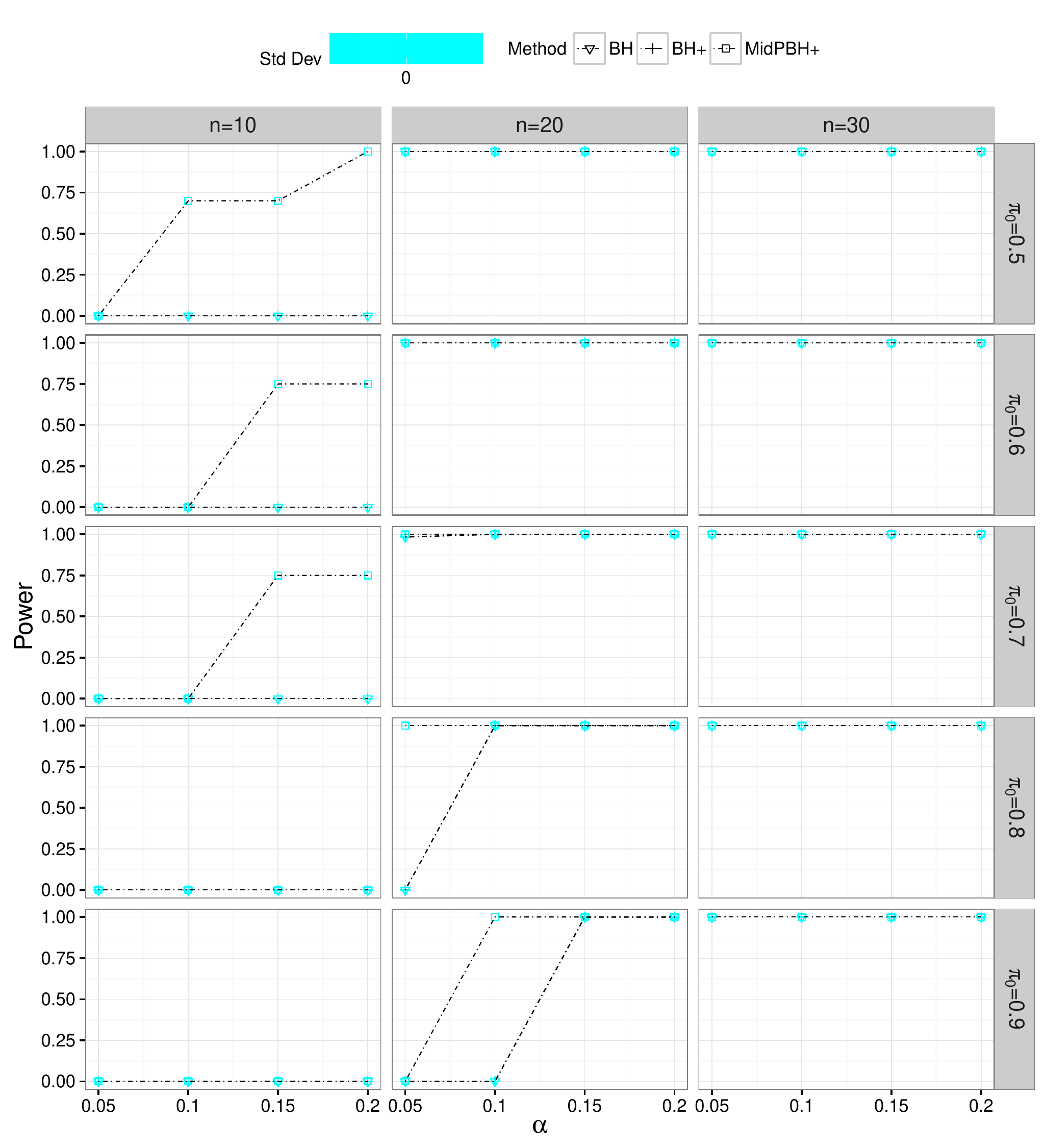}
\vspace{-0.5cm} \caption[Power for FET]{Power of each procedure. The color
legend ``Std Dev" denotes the standard deviation of the true discovery
proportion (TDP) whose expectation is power. Each procedure has been applied
to two-sided p-values of Fisher's exact tests under positive dependence.}%
\label{figPwrFETDep}%
\end{figure}

Now we summarize the results under dependence. For the simulation design
stated in \autoref{SecSimDesign}, the FDR of each procedure for each value of
the triple $\left(  \alpha,\pi_{0},\eta\right)  $ is zero, possibly due to the
positive dependence and approximately equal correlation between the generated
data and moderate signal sizes under the false null hypotheses. For BT's (see
\autoref{figPwrBTDep}), the powers of the procedures behave similarly to those
under independence but the improvements in power of MidPBH+ over BH+ seems to
be larger when Poisson random variables have smaller means. However, compared
to the independence case, such improvements seem to decrease quicker as the
Poisson means or the total number of trials for Binomial distributions
increases. In contrast, for FET's, the improvements in power of MidPBH+ over
BH+ is either zero or enormous, likely due to the dependence among the data;
see \autoref{figPwrFETDep}.

\section{Three applications of the BH+ procedure}

\label{seAppAll}

We provide three applications of the BH+ procedure to multiple testing based
on discrete and heterogeneous p-value distributions: one in a differential
methylation study on Arabidopsis thaliana, another an HIV study, and the other
a clinical safety study. BH+ applied to both conventional p-values and mid p-values, will be compared to BH
applied to conventional p-values. All procedures are
implemented at nominal FDR level $0.05$. The naming convention for each
procedure is the same as that in \autoref{secSimRes}.

\subsection{Application to methylation study}

\label{secAppMethy}

The aim of the study is to identity differentially methylated cytosines
between two unreplicated lines of Arabidopsis thaliana, wild-type (Col-0) and
mutant defective (Met1-3). Corresponding to each cytosine, the null hypothesis
is \textquotedblleft the cytosine is not differentially methylated between the
two lines". The data set is available from \cite{Lister:2008}.
There are $22265$ cytosines in each line, and each cytosine in each line has a
discrete count that indicates its level of methylation. We choose cytosines
whose total counts for both lines are greater than $10$ and whose count for
each line does not exceed $25$, in order to filter out genes with unreliable
low counts and to better utilize for multiple testing the jumps in the p-value
distributions. This yields $2785$ cytosines, i.e., $2785$ null hypotheses to
test simultaneously.

We model the counts for each cytosine in the two lines by
two independent Poisson distributions, and use Binomial test to test each null
hypothesis. MidPBH+ makes $531$ discoveries whereas each of BH and BH+ makes $420$, illustrating
the power improvement by BH+ based on mid p-values.

\subsection{Application to HIV study}

\label{secAppHIV}

The aim of the study is to identify, among $m=118$ positions, the
\textquotedblleft differentially polymorphic\textquotedblright\ positions,
i.e., positions where the probability of a non-consensus amino-acid differs
between two sequence sets. The two sequence sets were obtained from $n=73$
individuals infected with subtype C HIV (categorized into Group 1) and $n=73$
individuals with subtype B HIV (categorized into Group 2), respectively. The
data set is available from \cite{Gilbert:2005}, and how multiple testing is
set up based on p-values of FET's can also be found there. In summary, each
position on the two sequence sets corresponds to a null hypothesis that
\textquotedblleft the probabilities of a non-consensus amino-acid at this
position are the same between the two sequence sets\textquotedblright.

There are $50$ positions for which the total observed counts are identically
$1$ and the corresponding two-sided p-value CDF's are Dirac masses. To reduce
the uncertainty induced by positions whose observed total counts are too low,
we only analyze those whose observed total counts are at least $5$. This gives
$41$ positions, i.e., $41$ null hypotheses to test. MidPBH+ makes $25$ discoveries, considerably more
than $16$ made by each of BH and BH+.

\subsection{Application to clinical safety study}

The study aimed at differentiating adverse experiences that might have been
caused by a vaccine. The study design, detailed by \cite{Mehrotra:2004}, can
be summarized as follows. Participants were randomly assigned to receive the
quadrivalent measles, mumps, rubella and varicella (MMRV) on day 0 (Group 1
with $148$ toddlers) or the trivalent MMR on day 0 followed by varicella on
day 42 (Group 2 with $132$ toddlers). The safety profile of a vaccine is
represented by reported cases for each of $40$ adverse experiences. Table 1 of
\cite{Mehrotra:2004} shows the safety profile of MMRV and that of varicella
alone that were recorded for Group 1, days 0 to 42, and Group 2, days 42 to 84.

For each adverse experience, the null hypothesis is that \textquotedblleft
varicella is not associated with the adverse experience\textquotedblright, and
FET is conducted to test if the probabilities of the adverse experience are
the same between the two groups by assuming that the numbers of reported cases
are realizations of two independent Binomial distributions with total numbers
of trials $148$ and $132$ respectively. In this application, none of BH, BH+ and MidPBH+ claimed that
varicella is not associated with any of the $40$ adverse experiences.

\section{Discussion}

\label{secDisc}

We have proposed the BH+ procedure for FDR\ control for multiple testing based
on p-values with c\`{a}dl\`{a}g distribution functions. BH+ generalizes the BH
procedure and is at least as powerful as the latter when they are applied to
super-uniform p-values. Further, it is usually more powerful when applied to
mid p-values than when applied to conventional p-values, and can be much so. A
theoretical justification for this has been provided. Our work opens the door
for multiple testing based on p-values with general distributions. Further,
the BH+ procedure can be extended into weighted FDR procedures and procedures
for multilayer FDR control based on p-values with c\`{a}dl\`{a}g distribution
functions. We leave the investigation of these to future research.

\section*{Acknowledgements}

I would like to thank Sanat K. Sarkar for discussions on the BH+ procedure, Sebastian D\"{o}hler and Ruther Heller
for comments on the FDR behavior of multiple testing based on mid p-values,
and Joseph F. Heyse for sharing the clinical safety data.

%

\bibliographystyle{dcu}


\clearpage
\appendix{}

\section*{Appendices}

We provide in \autoref{secProof} proofs of \autoref{lmGeneral},
\autoref{ThmMidVersusCon} and in
\autoref{secSimDep} simulation results under dependence.

\section{Proofs}

\label{secProof}

\subsection{Proof of \autoref{lmGeneral}}

Let $\hat{\alpha}$ be the FDR\ of the BH+ procedure at nominal FDR level
$\alpha\in\left(  0,1\right)  $. Let $b_{i,r}=\Pr\left(  \left.
R=r\right\vert p_{i}\leq\gamma_{r}\right)  $ for $i\in I_{0}$ and
$r\in\left\{  1,...,m+1\right\}  $ for which $b_{i,m+1}=0$ is set for each
$i\in I_{0}$. Then the definitions of $F^{\ast}$ and $\gamma_{r}$ imply%
\begin{equation}
\hat{\alpha}=\sum_{i\in I_{0}}\sum_{r=1}^{m}\frac{1}{r}\Pr\left(  p_{i}%
\leq\gamma_{r}\right)  b_{i,r}\leq\sum_{i\in I_{0}}\sum_{r=1}^{m}\frac{1}%
{r}F^{\ast}\left(  \gamma_{r}\right)  b_{i,r}\leq\sum_{i\in I_{0}}\sum
_{r=1}^{m}\frac{\alpha}{m}b_{i,r}. \label{eqc72}%
\end{equation}
Let \thinspace$e_{i}\left(  s,t\right)  =\Pr\left(  \left.  R\geq s\right\vert
p_{i}\leq t\right)  $ for $i\in I_{0}$, $s\in\left\{  1,...,m+1\right\}  $ and
$t\in\left[  0,1\right]  $ for which $e_{i}\left(  m+1,t\right)  =0$ is set
for each $i\in I_{0}$ and $t\in\left[  0,1\right]  $. Then $b_{i,r}%
=e_{i}\left(  r,\gamma_{r}\right)  -e_{i}\left(  r+1,\gamma_{r}\right)  $ for
each $i\in I_{0}$ and $1\leq r\leq m$. Further, the non-decreasing property of
$\gamma_{r}$ in $r$ and the PRDS property of the p-values imply%
\begin{equation}
b_{i,r}=e_{i}\left(  r,\gamma_{r}\right)  -e_{i}\left(  r+1,\gamma_{r}\right)
\leq e_{i}\left(  r,\gamma_{r}\right)  -e_{i}\left(  r+1,\gamma_{r+1}\right)
. \label{eqc3}%
\end{equation}
So,
\begin{equation}
\sum_{i\in I_{0}}\sum_{r=1}^{m}\frac{\alpha}{m}b_{i,r}\leq\frac{\alpha}{m}%
\sum_{i\in I_{0}}\sum_{r=1}^{m}\left[  e_{i}\left(  r,\gamma_{r}\right)
-e_{i}\left(  r+1,\gamma_{r+1}\right)  \right]  \leq\frac{m_{0}\alpha}{m}%
\leq\alpha. \label{eqc5}%
\end{equation}
Combining (\ref{eqc5}) with (\ref{eqc72}) justifies $\hat{\alpha}\leq\alpha$.

Now we show the second claim. When $\left\{  p_{i}\right\}  _{i=1}^{m}$ are
super-uniform, $F_{\ast}\left(  t\right)  \leq t$ for each $t\in\left[
0,1\right]  $ and $\gamma_{k}\geq\frac{\alpha k}{m}$ for each $1\leq k\leq m$.
Since the BH procedure is a step-up procedure with critical values $\left\{
\frac{\alpha k}{m}\right\}  _{k=1}^{m}$ and rejects each $H_{i}$ if its
associated p-value $p_{i}\leq\frac{\alpha R^{\prime}}{m}$ whenever%
\[
R^{\prime}=\max\left\{  1\leq i\leq m:p_{\left(  k\right)  }\leq\frac{\alpha
k}{m}\right\}
\]
is well defined, we see that $R^{\prime}$ is upper bounded by $R$ almost
surely. Namely, the set of null hypotheses rejected by the BH procedure is
almost surely contained in that of the BH+ procedure. This completes the proof.

\subsection{Proof of \autoref{ThmMidVersusCon}}

Recall from \autoref{secDefinePval} that $\Pr\left(  P\left(  X\right)  \leq
P\left(  x\right)  \right)  =P\left(  x\right)  $ for all $x\in\mathcal{S}$
and that $\Pr\left(  Q\left(  X\right)  \leq Q\left(  x\right)  \right)
=P\left(  x\right)  $ for all $x\in\mathcal{S}$. Then, $\Pr\left(  P\left(
X\right)  \leq t\right)  \leq\Pr\left(  Q\left(  X\right)  \leq t\right)  $
for each $t\in\left[  0,1\right]  $. So, $F_{i}^{\mathsf{mp}}\geq
F_{i}^{\mathsf{cp}}$ for each $i=1,\ldots,m$. However, $W_{\mathsf{mp}}%
=\max_{1\leq i\leq m}F_{i}^{\mathsf{mp}}$ and $W_{\mathsf{cp}}=\max_{1\leq
i\leq m}F_{i}^{\mathsf{cp}}$. Thus, $W_{\mathsf{mp}}\geq W_{\mathsf{cp}}$.
This justifies the first claim.

Let $S_{\mathsf{cp}}$ and $S_{\mathsf{mp}}$ be the support of $W_{\mathsf{cp}%
}$ and $W_{\mathsf{mp}}$ respectively. For each $k=1,\ldots,m$, let%
\[
\gamma_{k}^{\mathsf{cp}}=\max\left\{  t\in S_{\mathsf{cp}}:W_{\mathsf{cp}%
}\left(  t\right)  \leq\frac{\alpha k}{m}\right\}
\]
and%
\[
\gamma_{k}^{\mathsf{mp}}=\max\left\{  t\in S_{\mathsf{mp}}:W_{\mathsf{mp}%
}\left(  t\right)  \leq\frac{\alpha k}{m}\right\}  .
\]
Recall $\left\{  P_{\left(  i\right)  }\right\}  _{i=1}^{m}$ as the order
statistics of $\left\{  P_{i}\right\}  _{i=1}^{m}$ and $\left\{  Q_{\left(
i\right)  }\right\}  _{i=1}^{m}$ those of $\left\{  Q_{i}\right\}  _{i=1}^{m}%
$. Then $R_{\mathsf{cp}}=\max\left\{  i:P_{\left(  i\right)  }\leq\gamma
_{i}^{\mathsf{cp}}\right\}  $ and $R_{\mathsf{mp}}=\max\left\{  i:Q_{\left(
i\right)  }\leq\gamma_{i}^{\mathsf{mp}}\right\}  $, where $R_{\mathsf{cp}}=0$
or $R_{\mathsf{mp}}=0$ is set if the set $\left\{  i:P_{\left(  i\right)
}\leq\gamma_{i}^{\mathsf{cp}}\right\}  $ or $\left\{  i:Q_{\left(  i\right)
}\leq\gamma_{i}^{\mathsf{mp}}\right\}  $ is empty, respectively.

If $R_{\mathsf{cp}}=0$, then the BH+ procedure based on $\left\{
P_{i}\right\}  _{i=1}^{m}$ makes no rejections, and $R_{\mathsf{mp}}\geq
R_{\mathsf{cp}}$ holds automatically. It is left to consider the case where
$R_{\mathsf{cp}}$ is between $1$ and $m$. Suppose (\ref{eq15a}) holds, i.e.,
$W_{\mathsf{mp}}\left(  Q_{\left(  R_{\mathsf{cp}}\right)  }\right)
\leq\alpha m^{-1}R_{\mathsf{cp}}$. Then $Q_{\left(  R_{\mathsf{cp}}\right)
}\leq\gamma_{R_{\mathsf{cp}}}^{\mathsf{mp}}$, $R_{\mathsf{mp}}$ is well
defined, and $R_{\mathsf{mp}}\geq R_{\mathsf{cp}}$. On the other hand, if
$R_{\mathsf{mp}}\geq R_{\mathsf{cp}}$, then $Q_{\left(  R_{\mathsf{cp}%
}\right)  }\leq\gamma_{R_{\mathsf{cp}}}^{\mathsf{mp}}$ and (\ref{eq15a})
holds. This justifies the second claim and completes the proof.

\newpage

\section{Simulation results under positive dependence}

\label{secSimDep}

\end{document}